\documentclass[lettersize,journal]{IEEEtran}
\usepackage{amsmath,amsfonts}
\usepackage{algorithmic}
\usepackage{array}
\usepackage[caption=false,font=normalsize,labelfont=sf,textfont=sf]{subfig}
\usepackage{textcomp}
\usepackage{stfloats}
\usepackage{url}
\usepackage{verbatim}
\usepackage{graphicx}
\usepackage[hypertexnames=false]{hyperref}
\usepackage{booktabs}
\usepackage{makecell}
\usepackage{biblatex}
\def\BibTeX{{\rm B\kern-.05em{\sc i\kern-.025em b}\kern-.08em
    T\kern-.1667em\lower.7ex\hbox{E}\kern-.125emX}}
\usepackage{balance}

\begin{document}
\title{Predictive Coding Light+: learning to predict visual sequences with spike timing-dependent plasticity and synaptic delays}
\author{Antony W. N'dri$^{1,2}$, Thomas Barbier$^{1}$, Céline Teulière$^{1,*}$, Jochen Triesch$^{1,3,4,*,\dagger}$\\
$^{1}$ Universit\'{e} Clermont Auvergne, Clermont Auvergne INP, CNRS, Institut Pascal,\\ F-63000 Clermont-Ferrand, France\\
$^{2}$ Orange Labs, Sophia Antipolis, France\\
$^{3}$ Frankfurt Institute for Advanced Studies,
Frankfurt am Main, Germany\\
$^{4}$ Goethe University Frankfurt, Institute of Computer Science, Frankfurt am Main, Germany\\
$^{*}$ Both authors jointly supervised this work.\\
$^{\dagger}$ Corresponding author: {\tt\small {\href{mailto:triesch@fias.uni-frankfurt.de}{triesch@fias.uni-frankfurt.de} }}
}

\maketitle


\begin{abstract}
\textbf{The ability to predict the future is of great value for biological and artificial cognitive systems alike.
However, successfully predicting the future typically requires maintaining a memory of the recent past. It is currently unclear how biological or artificial spiking neural networks can learn to maintain past sensory information to help predict the future. Here we propose Predictive Coding Light+ (PCL+), a spiking neural network architecture for unsupervised sequence processing that learns recurrent excitatory connections with delays to enable short-term retention of information. We show that the PCL+ network reproduces classic findings on sequence learning in visual cortex. Furthermore, it learns to ``fill in'' missing input in a challenging gesture recognition task. Overall, our work shows how spiking neural networks can learn recurrent excitatory connections with delays to maintain a record of the recent past and successfully predict the future.}
\end{abstract}


\section{Introduction}


\begin{figure*}[ht]
	\centering
	\includegraphics[]{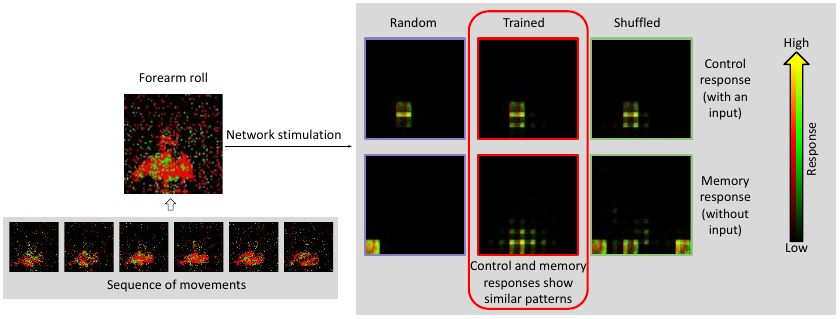}
	\caption[A PCL+ network ``fills in'' missing sensory input.]{\textbf{A PCL+ network ``fills in'' missing sensory input.} Top left: event-based input pattern representing a forearm roll gesture. Bottom left: example movement sequence. Right: example of retinotopically organized average simple cell responses for two input conditions (top and bottom row) and three network architectures (columns). The two input conditions are a network stimulated with visual input (top row) and a network where the later part of an input sequence is omitted (bottom row). We consider (left to right)  a PCL+ network with random recurrent excitatory connectivity (Random), a trained PCL+ network (Trained), and a trained PCL+ network where recurrent excitatory connections were shuffled after training (Shuffled). Image regions triggering low/high activity in the network are represented by dark/bright colors and the contrast of the images was increased by 50\% for better visualization. The trained PCL+ network exhibits similar activity patterns when the visual input is present and when part of it is omitted, thereby demonstrating the biologically observed ``filling in'' of predictable inputs that are omitted.}
	\label{fig:figure1}
\end{figure*}


The ability to predict the future is a hallmark of both biological and artificial cognitive system. Usually, to successfully predict the future it is helpful to keep track of what has happened in the recent past. This can be realized by various kinds of short-term or working memory architectures. For example, in deep learning systems, working memory functionality is often realized using recurrent neural networks \cite{ schuster1997bidirectional,sherstinsky2020fundamentals} or large context windows of transformer architectures \cite{vaswani2017attention, han2022survey}.


In contrast to such deep learning architectures, biological neurons encode and transmit information using discrete voltage pulses called ``spikes'' \cite{gerstner2002spiking}.
Spiking neural networks \cite{tavanaei2019deep, yamazaki2022spiking} mimic biological neurons more accurately have been shown to consume little energy when implemented on neuromorphic hardware \cite{davies2021advancing, gonzalez2024spinnaker2}.

How biological networks of such spiking neurons retain encoded information for short amounts of time is an unresolved question \cite{barak2014working}. Traditionally, working memory has been associated with persistent activity \cite{curtis2003persistent, barak2014working}. Fundamentally, networks of spiking neurons can exhibit persistent activity through the usage of recurrent excitatory synaptic connections. However, building recurrent excitatory networks remains challenging as they may suffer from runaway activity \cite{turrigiano2004homeostatic, miehl2022stability, zenke2017temporal}. 

Spiking is also known to be a major source of energy consumption in the brain \cite{lennie2003cost}. If short-term memory traces are implemented through delay activity in recurrent spiking networks, this consumes precious energy, suggesting a trade-off between memory capacity and energy efficiency. However, this aspect is not addressed in spiking models of working memory based on, e.g., dynamic neural fields \cite{sandamirskaya2014dynamic, de2016event, evanusa2019event, kreiser2018organizing} and liquid state machines \cite{maass2002real, maass2011liquid}.

Theories of brain function such as predictive coding \cite{rao1999predictive} argue that neural populations predict their inputs so that they only need to send  prediction errors to higher processing areas, saving precious energy. A memory trace is perfectly predictable from the past, however. In fact, otherwise it would not be a very faithful memory. So mechanisms of short-term memory storage and the associated energetic costs are typically not considered.

Here, we bring together the perspectives of working memory and energy efficiency by introducing {\em Predictive Coding Light+} (PCL+). PCL+ is an extension of the PCL network \cite{ndri2023, n2025predictive} that adds recurrent excitation with synaptic delays to provide working memory functionality. This enables it to function as an associative memory for temporal sequences. In the original PCL network, neurons suppress the most predictable spikes through adaptive recurrent and feedback inhibition. This enables the PCL network to reproduce various biological observations considered to be signatures of an efficient neural code \cite{ndri2023}. PCL+ keeps these functionalities and further incorporates an adaptive working memory using heterogeneous synaptic delays \cite{izhikevich2006polychronization}. In PCL+, while neurons can remove predictable spikes, they are also given the ability to retain past information in the neural code through delayed feedback excitatory pathways that take the form of lateral and top-down synaptic excitatory connections. These connections also learn via standard spike timing-dependent learning ruels. Each spike in a PCL+ network is thus also sent out into the ``future'' using those delayed excitatory connections. Inhibition instead occurs instantaneously and removes immediately redundant spikes, thus ensuring efficiency. Thus, PCL+ harnesses the simple and elegant concept of instantiating a working memory in a spiking neural networks by propagating spikes through time using synaptic delays.


We test PCL+ by learning spatiotemporal sequences on inputs from an event-based vision sensor. When trained on sequences of gratings, PCL+ reproduces experiments on visual sequence learning in mouse primary visual cortex \cite{gavornik2014learned}. Furthermore, when applied to videos of gestures, PCL+ predicts future inputs enabling it to ``fill in'' missing information. This demonstrates that it can function as an associative memory for temporal sequences (see Fig.~\ref{fig:figure1}). Overall, PCL+ proposes a simple architecture to combine energy-efficient information processing and working memory functionality without spending extra spikes.

\section{Methods}

\subsection{Network architecture}

\begin{figure*}[ht]
	\centering
	\includegraphics[]{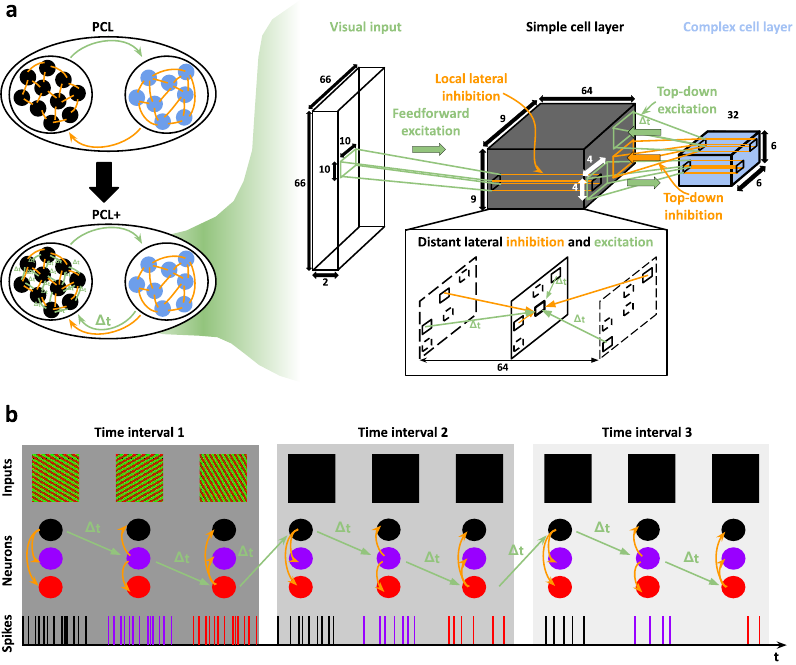}
	\caption{\textbf{Predictive Coding Light+. a,} Left: Simplified architecture of the PCL and PCL+ networks illustrating excitatory (green) and inhibitory (ornage) connections. The PCL+ network contains additional recurrent excitatory connections with conduction delays (green, $\Delta t$). Right: detailed architecture of the PCL+ network with all types of inhibition: local lateral, distant lateral, top-down inhibitions and the different types of excitation: feedforward, distant lateral and top-down excitations. Distant lateral and top-down excitation both use synaptic delays whereas all other synaptic connections are instantaneous. \textbf{b,} Concept behind the delayed recurrent excitation in a PCL+ network. In time interval 1, when neurons are excited with a sequence of stimuli, they learn to associate the different stimulus representations using recurrent excitation. When the stimulus disappears in time intervals 2 and 3, a memory of the previous stimulus sequence is maintained in the network through the delayed recurrent excitation, but it decays due to recurrent inhibition, illustrating a fading memory. }
	\label{fig:figure2}
\end{figure*}

Predictive Coding Light+ (PCL+) extends the PCL model \cite{n2025predictive} by incorporating recurrent excitatory connections with synaptic delays to facilitate the short-term retention of information. Figure~\ref{fig:figure2} illustrates the PCL+ network architecture and its differences from the original PCL network. We will first describe the original PCL network and then detail the extensions of PCL+.

\begin{table*}[ht!]
	\caption[Connectivity parameters of the PCL+ network.]{Connectivity parameters of the PCL+ network. Connectivity parameters values indicate dimensions of the following form: $(x \times y \times z)$ cells. }
	\label{tab:PCL+architecture}
	\footnotesize
	\centering
	\begin{center} \textit{Simple cells' connectivity parameters:} \end{center}
	\resizebox{0.75\columnwidth}{!}{
		\begin{tabular}{ l l l }
			\toprule 
			\thead{Retina size} & \thead{Input zone} & \thead{$\rm N_{cells}$} \\
			\midrule
			\thead{$(346 \times 260 \times 2)$} & \thead{$(66 \times 66 \times 2)$}  & \thead{$(9 \times 9 \times 64)$}\\
			\midrule
			\thead{Receptive field} &  \thead{Overlap} &\thead{Plastic lateral inhibition and excitation range}\\ 
			\midrule
			\thead{$(10 \times 10 \times 2)$} & \thead{$(3 \times 3 \times 2)$} & \thead{$(4 \times 4 \times 64)$}\\ 
			\bottomrule
	\end{tabular}}
	\begin{center} \textit{Complex cells' connectivity parameters:} \end{center}
	\resizebox{0.75\columnwidth}{!}{
		\begin{tabular}{ l l l  }
			\toprule
			\thead{Input size} & \thead{Input zone} & \thead{$\rm N_{cells}$} \\
			\midrule
			\thead{$(9 \times 9 \times 64)$} &  \thead{$(9 \times 9 \times 64)$} & \thead{$(6 \times 6 \times 32)$} \\
			\midrule
			\thead{Receptive field} &  \thead{Overlap} &\thead{Plastic top-down inhibition and excitation range}\\ 
			\midrule
			\thead{$(4 \times 4 \times 64)$} & \thead{$(3 \times 3 \times 64)$} & \thead{$(4 \times 4 \times 64)$}\\
			\bottomrule \\
	\end{tabular}}
\end{table*}

\begin{table*}[ht]
	\caption[Neurons' membrane potential and plasticity parameters of the PCL+ network.]{Neurons' membrane potential and plasticity parameters.}
	\label{tab:PCL+neuronparams}
	\footnotesize
	\centering
	\begin{tabular}{cccccc}
		\toprule
		\multicolumn{6}{c}{Simple cells}                   \\
		\midrule
		\thead{$\rm V_{reset}$\\(mV)}  & \thead{$\rm V_{thresh}$\\(mV)} & \thead{$\rm \tau_{m/I}$\\(ms)} & \thead{$\rm \tau_{RP}$\\(ms)} & \thead{$\rm \tau_{LTP}$\\(ms)}  & \thead{$\rm \tau_{LTD}$\\(ms)} \\
		\midrule
		-10  & 30 & 18 & 5 & 7 & 7 \\
		\midrule 
		\thead{$\rm \eta_{RP}$\\(mV)} & \thead{$\rm w_{min}$\\(mV)} & \thead{$\rm w_{max}$\\(mV)} & \thead{$\rm \eta_{LTP/LTD}$\\(mV)} &   $\rm \eta_{+/-}$ & $\lambda$ \\
		\midrule
		10 & 0 & \thead{f-f. excit: 3\\dist. inhib.: 30\\td. inhib.: 30\\loc. inhib.: 30\\lat. excit.: 30\\td. excit.: 30} & \thead{f-f. excit: $\pm$0.000204\\dist. inhib.: $\pm$2.652\\td. inhib.: $\pm$1.428\\loc. inhib.: $\pm$0.612\\lat. excit.: $\pm$3.06\\td. excit.: $\pm$1.02} & \thead{f-f. excit: 0.33\\dist. inhib.: 0.033\\td. inhib.: 0.033\\loc. inhib.: 0.033\\lat. excit.: 0.033\\td. excit.: 0.033} & \thead{f-f. excit: 50\\dist. inhib.: 6500\\td. inhib.: 3500\\loc. inhib.: 1500\\lat. excit.: 750\\td. excit.: 250} \\
		\bottomrule \\ 
		
		\multicolumn{6}{c}{Complex cells}                   \\
		\midrule
		\thead{$\rm V_{reset}$\\(mV)}  & \thead{$\rm V_{thresh}$\\(mV)} & \thead{$\rm \tau_{m}$\\(ms)} & \thead{$\rm \tau_{RP}$\\(ms)} & \thead{$\rm \tau_{LTP}$\\(ms)}  & \thead{$\rm \tau_{LTD}$\\(ms)} \\
		\midrule
		-10  & 3 & 50 & 5 & 40 & 40 \\
		\midrule 
		\thead{$\rm \eta_{RP}$\\(mV)} & \thead{$\rm w_{min}$\\(mV)} & \thead{$\rm w_{max}$\\(mV)} & \thead{$\rm \eta_{LTP/LTD}$\\(mV)} &   $\rm \eta_{+/-}$ & $\lambda$ \\
		\midrule
		5 & 0 & \thead{excit: 4\\loc. inhib.: 25} & \thead{excit: $\pm$0.08\\loc. inhib.: $\pm$0.048} & \thead{excit: 0.25\\loc. inhib.: 0.04\\} & \thead{excit: 1000\\loc. inhib.: 600} \\
		\bottomrule
	\end{tabular}
\end{table*}

\paragraph{Predictive Coding Light (PCL) and PCL+} The PCL network is a hierarchical spiking neural network that takes inspiration from simple and complex cells in mammalian V1. In \cite{n2025predictive}, it takes the form of a 2-layer network where the first layer models V1's simple cells and the second layer models V1's complex cells. All synaptic connections are learned using biologically plausible spike timing-dependent plasticity (STDP) rules.

The sensory input of the original PCL network comes from an event camera, which is a bio-inspired sensor that mimics retinal processing by outputting asynchronous events at each pixel whenever brightness changes go above or below an internal threshold. Events come in two flavors, either ``ON'' for brightness increases or ``OFF'' for brightness decreases. In the PCL network, simple cells are driven by such events while complex cells are activated by the simple cells. 
Neurons in the PCL network do not respect Dale's law and  can both excite and inhibit their targets. There are two types of synaptic connections in a PCL network: excitatory feedforward and inhibitory recurrent connections. The inhibitory recurrent synapses come in three forms: 1) local lateral inhibitory synapses present in both simple and complex cells which prevent units sharing the same receptive field location to spike at the same time, similar to a Winner-Take-All (WTA) mechanism\cite{gupta2009hebbian, barbier2021spike}, 2) distant lateral inhibitory synapses present only in simple cells which let them inhibit responses to predictable inputs in neighboring simple cells that do not share the same receptive field location, 3) top-down inhibitory synapses from complex cells to simple cells which allow complex cells to suppress predictable spikes in simple cells. Importantly, the distant lateral inhibitory synapses and the top-down inhibitory synapses remove highly redundant spikes from the neural code.

Inhibitory synapses and feedforward excitatory synapses in the original PCL network transmit signals without delays. 
PCL+ adds two excitatory synapse types to the network: 1) delayed distant lateral excitatory synapses between simple cells and 2) delayed top-down excitatory synapses from complex cells to simple cells. Both permit a delayed excitation of simple cells and therefore a short-term maintenance of information. By default, these delays are heterogeneous and drawn randomly from a wide uniform probability distribution [100~ms, 500~ms], unless specified otherwise. In some experiments, we also use identical delays for all recurrent excitation. Note that in contrast to other forms of memory the stored information is only accessible when a signal arrives at the target cell after the particular delay of that connection. We use the same connectivity parameters as in \cite{n2025predictive} for the connections that already existed in the PCL network. The parameters for the new excitatory connection types have been set to be the same as their inhibitory counterparts for simplicity. Please see Tab.~\ref{tab:PCL+architecture} for a summary of all connectivity parameters.

\subsection{Neuron model}
As its predecessor, PCL+ uses a leaky integrate-and-fire (LIF) neuron model. We set the resting membrane potential to 0~mV. When the membrane potential exceeds the threshold $V_{\theta}$, a neuron emits a spike and its membrane potential is reset to a value $V_{\rm reset} = -10~{\rm mV}$. After spiking, the neuron enters a relative refractory period in which it receives an additional hyperpolarizing current that prevents it from spiking again for a short period of time. Furthermore, to ensure that a neuron's membrane potential does not reach unrealistically low values, we set a minimum value $V_{\rm min}$ = -20~mV. Our network simulation is event-based: a neuron's membrane potential is only updated when it receives a synaptic input. Upon arrival of an excitatory or inhibitory synaptic input with weight $w_i$, a neuron's membrane potential is updated as:
\begin{equation}\label{eq1}
\tilde{V}(t+\Delta{}t) = \max \{ V_{\rm min}, V(t)e^{- \frac{\Delta{\rm t}}{\tau_{\rm m}}} + w_i(t) - \eta_{\rm RP}e^{-\frac{\rm t+\Delta{\rm t} - \rm t_{s}}{\tau_{\rm RP}}} \}\, ,
\end{equation}

\begin{equation}\label{eq2}
V(t+\Delta{}t) = 
\begin{cases}
\tilde{V}(t+\Delta{}t) & \text{ : } \tilde{V}(t+\Delta{}t) < V_\theta\\
V_{\rm reset} & \text{ : }\tilde{V}(t+\Delta{}t) \geq V_\theta \, ,
\end{cases}
\end{equation}
where $\Delta{\rm t}$ is the time since the last update of the membrane potential, $\tau_{\rm m}$ is the membrane time constant, $\eta_{\rm RP}$ is the maximum amplitude of the hyperpolarizing current giving rise to the relative refractory period, $\tau_{\rm RP}$ is the time constant of this current, and $t_{\rm s}$ is the time when the neuron last spiked. Please, see Tab.~\ref{tab:PCL+neuronparams} for the parameters of the neuron model as well as the parameters of the plasticity mechanisms.


\subsection{Synaptic plasticity mechanisms}

\subsubsection{Standard plasticity mechanisms inherited from PCL}
As a descendent of the PCL network \cite{n2025predictive}, PCL+ inherits most aspects of its plasticity mechanisms described in the following. 
\paragraph{Causal Spike Timing-Dependent Plasticity} We use a standard form of causal STDP to learn the feedforward excitatory weights and all of the inhibitory weights of our network. Synapses are strengthened when a pre-synaptic input arrived shortly before the spike (long term potentiation, i.e. LTP) and decreased when a pre-synaptic input arrived shortly after the spike (long term depression, i.e. LTD). As our simulations are event-based, we update the weights each time a neuron spikes, as follows:
\begin{equation}\label{eq4}
\tilde{\Delta{}}w_{i}^{\rm  LTP} = \eta_{\rm LTP} e^{\frac{t_{i} - t_{\rm s}}{\tau_{\rm LTP}}} \, ,
\end{equation}
\begin{equation}\label{eq5}
\tilde{\Delta{}}w_{i}^{\rm LTD} = -\eta_{\rm LTD} e^{\frac{t_{s-1} - t_i}{\tau_{\rm LTD}}} \, ,
\end{equation}
where $\eta_{\rm LTP}$ and $\eta_{\rm LTD}$ are the learning rates, respectively, for potentiation and depression, $t_{\rm i}$ is the spike time of the last input received from synapse $i$, $t_{\rm s}$ is the spike time of the post-synaptic neuron, $t_{s-1}$ is the spike time of the before-the-last spike of the post-synaptic neuron, $\tau_{\rm LTP}$ and $\tau_{\rm LTD}$ are time constants that control the temporal window of potentiation and depression, respectively.

\paragraph{Soft-bound synaptic plasticity}
We define a soft-bound synaptic plasticity mechanism that we use to reduce the risk of large weight changes. It keeps the weights between a maximum value $w_{\rm max}$ and a minimum value $w_{\rm min}$. We implement it using the two following functions:
\begin{equation}\label{eq6}
A_{+}(w_{i})= (w_{\rm max} - w_{i})\eta_{+} \, ,
\end{equation}
\begin{equation}\label{eq7}
A_{-}(w_{i})= (w_{i} - w_{\rm min})\eta_{-} \, ,
\end{equation}
where $\eta_{+}$ and $\eta_{-}$ are positive constants. The STDP updates then take the following form:
\begin{equation}\label{eq8}
\Delta{}w_{i}^{LTP} = A_{+}(w_{i})\tilde{\Delta{}}w_{i}^{LTP} \, ,
\end{equation}
\begin{equation}\label{eq9}
\Delta{}w_{i}^{LTD} = A_{-}(w_{i})\tilde{\Delta{}}w_{i}^{LTD} \, ,
\end{equation}

\paragraph{Weight normalization}

We also implement a weight normalization mechanism for all of our synaptic weights to control the total strength of the excitatory and inhibitory inputs that a neuron can receive. This effect mimics the competition over a limited number of resources that is thought to take place in biological brains \cite{triesch2018competition}. After each synaptic weight update (LTP and LTD), the weights are then normalized following:
\begin{equation}\label{eq10}
w_i \leftarrow \lambda \frac{w_i}{\sum_i w_i} \; \forall i  \, ,
\end{equation}
where $\lambda$ is the normalization strength target for incoming synapses received by a post-synaptic neuron. 

\paragraph{Weight sharing}
We use a weight sharing mechanism for our feedforward excitatory connections and our local lateral inhibitory connections for both simple and complex cells. While such a mechanism is not biologically plausible, it increases training speed and reduces memory requirements of the simulations. After a given synaptic weight has been updated and the subsequent weight normalization, the corresponding excitatory or local lateral inhibitory synapses at other retinal locations are set to the same values:
\begin{equation}\label{eq11}
\mathbf{\forall z, \forall j \neq k: w^{z, j}_i \leftarrow w^{z, k}_i \, ,}
\end{equation}
where $z$ gives the feature channel, $j$ gives the retinal location of the neuron, $k$ is the retinal location of the neuron that received the last synaptic weight update and $w_{i}$ denotes an excitatory or local lateral inhibitory synaptic weight indexed by $i$.

\subsubsection{Activity stabilization mechanisms in PCL+}
To mediate stability issues due to its recurrent excitatory connectivity, PCL+ uses additional mechanisms beyond the standard mechanisms of the PCL network. These operate at two different stages, learning and inference. 

\paragraph{Plasticity regulation}
Recurrent excitation significantly increases the number of spikes and tends to lead to issues when learning with STDP: if two neurons make each other spike, their weights increase from LTP which leads to further spikes and then further weight increases to the point that networks' spikes are no longer encoding the sensory input. We alleviate this problem by incorporating a learning rule which regulates LTP and LTD according to feedforward and recurrent excitatory synaptic input levels. We define the trace $I$ of a synaptic input as: 

\begin{equation}\label{eq12}
I(t + \Delta t) = I(t) e^{- \frac{\Delta t}{\tau_{\rm I}}} + w_{\rm i}(t) \, ,
\end{equation}
where $\Delta t$ is the time since the last update of the membrane potential, $\tau_{\rm I}$ controls the decay of the trace and $w_{\rm i}(t)$ is a synaptic input. We denote as $I_{\rm ffExc}$ and $I_{\rm rExc}$, respectively, the traces of feedforward and recurrent excitatory synaptic inputs received by a neuron. 

Second, we define a symmetric STDP update rule in which only LTP occurs:
\begin{equation}\label{eq13}
\tilde{\Delta{}}w_{i}^{\rm  LTP,symmetric} = \eta_{\rm LTP} \left(e^{\frac{t_{i} - t_{\rm s}}{\tau_{\rm LTP}}} + e^{\frac{t_{s-1} - t_{\rm i}}{\tau_{\rm LTP}}}\right) \, ,
\end{equation}
\begin{equation}\label{eq14}
\Delta{}w_{i}^{LTP, symmetric} = A_{+}(w_{i})\tilde{\Delta{}}w_{i}^{LTP, symmetric} \, ,
\end{equation}
where all parameters stay the same. 

Next, we pair the synaptic input trace with the above symmetric STDP rule such that plasticity follows: 
\begin{equation}\label{eq15}
\tilde \Delta{}w_{\rm i}^{\rm  rExc, s} =  \begin{cases}
\log \left(  \frac{I_{\rm ffExc}}{\beta \times I_{\rm rExc}} \right) {\Delta{}w}_{\rm i}^{\rm  LTP, symmetric} & \text{if $I_{\rm rExc} > 0$}\\
0 & \text{if $I_{\rm rExc} = 0$,}
\end{cases}
\end{equation}
where $\tilde \Delta{}w_{\rm i}^{\rm  rExc, s}$ refers to the synaptic weight update of simple cells' recurrent excitatory weights and $\beta$ is a parameter that controls the influence of the ratio of excitatory inputs. We set $\beta$ to 1. 

Following the above rule, whenever a spike occurs and a recurrent excitatory synapse is updated, if feedforward inputs are stronger than recurrent inputs, the weight gets potentiated since the logarithmic term would be positive. However, if feedforward inputs are weaker, the rule gets reversed and switches to LTD as the logarithmic term becomes negative. In particular, in the absence of feedforward activity, recurrent excitatory connections are reduced which prevents runaway dynamics of recurrent excitation. 

Finally, we bound the weight updates to avoid to large increases (or decreases). Our final recurrent excitatory weight update rule follows:
\begin{equation}\label{eq16}
\Delta{}w_{\rm i}^{\rm  rExc, s} =  \begin{cases}
\tilde \Delta{}w_{\rm i}^{\rm  rExc, s} & \text{if } -2 \leq \tilde \Delta{}w_{\rm i}^{\rm  rExc, s} \leq 2,  \\
2 & \text{if } \tilde \Delta{}w_{\rm i}^{\rm  rExc, s} \geq  2 , \\
- 2 & \text{if } \tilde \Delta{}w_{\rm i}^{\rm  rExc, s} \leq - 2 \text{ and } I_{ffExc} = 0,  \\
\end{cases}
\end{equation}
Overall, our plasticity modulation mechanism aims at stabilizing network activity while learning to retain information across delays. 

\paragraph{Inhibition regulation}
Neurons in the PCL+ network also regulate the amount of inhibition they receive. Each PCL+ neuron is driven by feedforward and recurrent excitatory inputs. However, each time a spike happens, inhibitory inputs are generated from the diverse set of inhibitory synapses. While feedforward inputs from event cameras tend to be strong and require a strong amount of inhibition for the removal of predictable spikes, recurrent excitatory inputs are weaker and would be fully canceled  if the amount of inhibition attributed to the two input types were the same. To prevent this, we modulate distant lateral and top-down inhibitory currents so that their strengths depend on the amount of feedforward and recurrent excitatory inputs. We reduce inhibition when a neuron spikes despite little feedforward excitation, such that its response is amplified. We motivate such a mechanism by biological observations of disinhibition during sensory experience that suggest its importance in learning and the formation of memory \cite{letzkus2015disinhibition}. Let $\hat{w}_{\rm i}^{rI}$ be the modulated amount of distant lateral or top-down inhibitory input. It evolves according to:

\begin{equation}
\hat{w}_{\rm i}^{rI} = {w}_{\rm i}^{rI} e^{-\frac{I_{\rm rExc}}{I_{\rm ffExc} + \alpha \times I_{\rm rExc}}} \, ,
\end{equation}

where $\alpha$ is a constant that we set here to 0.5. When a neuron receives little to no excitation (both feedforward and recurrent), it also receives little to no inhibition. This allows a network to enter directly a ``memory'' or ``recall'' mode whenever no outside input is received. 

\subsection{Experiments}
\subsubsection{Datasets}
\paragraph{Natural images} In \cite{n2025predictive}, an event-based version of the Van Hateren natural images database \cite{van1998independent} was proposed. It consists of 2000 event-based samples made from converting individual images into 800~ms long samples by moving the image into 16 random directions that cover the scope of 360°. This dataset contains various shots that would be seen during a natural vision experience, such as: forests, lakes, buildings, cars, humans, flowers, grass and so on. We use this dataset to first train the feedforward excitatory and recurrent inhibitory connections of the PCL+ network. 

\paragraph{Gratings} We generate event-based sequences of gratings using the PIX2NVS \cite{bi2017pix2nvs} event-based simulator. This simulator converts frames to events by calculating the logarithmic difference between pixel values of successive frames and emitting a random number of events (that we set here between 1 and 5) only when this difference is superior to a threshold that is updated individually for each pixel. While there are other simulators (e.g. the v2e simulator \cite{hu2021v2e}) that implement more finely the different characteristics of event-based vision sensors, PIX2NVS was sufficient in our case since we consider simple stimuli and movements. Essentially, we use this simulator to generate 250~ms sinusoidal gratings of various orientations (0°, 30°, 60°, 120° and 150°) with a phase of 0°, a spatial frequency of 1/6 cycles per degree and a temporal frequency of 4~Hz. We construct sequences of such grating stimuli to train our recurrent excitation and inhibition. Please see Fig.~\ref{fig:figure3}a for an example of such a sequence of gratings. Using these sequences of gratings, we train our recurrent excitatory connections and further fine-tune our distant lateral and top-down inhibitory connections to learn to recognize and predict spatiotemporal sequences as observed in mouse V1 \cite{gavornik2014learned}.

\paragraph{DVS128 Gesture} This dataset was proposed in \cite{amir2017low}. It consists of a set of 11 classes of gestures, each performed by 29 individuals under 3 different illuminations. In this work, similarly to the original PCL work, we scale down the spatial extent by a factor of 1.94 so that the gestures match the size of our network's input zone. In this dataset, the gestures of the first 23 individuals are used for training and the remaining ones are used for testing. We use this dataset to evaluate the ability of our network to learn complex spatiotemporal sequences. Specifically, this evaluation is done via two tasks: time-series forecasting, and classification that we describe next.

\subsubsection{Spike train descriptor}
To evaluate time-series forecasting as well as to perform classification with PCL+, we build a feature descriptor using the activity of our simple cells and complex cells. First, we divide all of our sequences into a number of time bins $B$ and for each time bin, we count the number of spikes for individual cells. We then combine these values into a vector of size  $B \times N$ where $N$ is the number of cells (number of simple cells, number of complex cells, or both depending on the case studied). We use bin widths of size 1.5~s.


\subsubsection{Time-series forecasting} 
We evaluate PCL+ on a time-series (here gestures) forecasting task. After training our recurrent excitatory, distant lateral and top-down inhibitory connections on the training database of gestures (only the first 3.5~s of each gesture), we test our network on its ability to predict the next simple and complex cells' spikes after being excited by a cue. The cue consists in a short presentation of the gesture which we choose here to be either 0.5~s or 1.5~s long. We then record the spontaneous activity of both simple and complex cells for 1.5~s by using our spike trains descriptor. As gestures represent a visual time-series that is here encoded in the network's spikes, this spontaneous activity is a prediction of the future activity of the network. To evaluate the quality of this prediction, we calculate the cosine similarity of the actual neural activity when gestures are fully presented, and this predicted neural activity in the absence of gestures. 


\section{Results}

\subsection{Training of the baseline network}
The PCL+ network is built on top of an already trained PCL network that has been shown to exhibit simple and complex cell-like receptive fields (See right of Fig.~\ref{fig:suppl1}b and Fig.~\ref{fig:suppl2}). This baseline network is trained for 5 epochs on the entire set of natural images with each sample being shown twice to the network, for a final sample duration of 1.6~s (instead of 800~ms). During this training, feedforward excitatory connections, local lateral, distant lateral and top-down inhibitory connections are all trained simultaneously. In \cite{gavornik2014learned}, neural response increases and ``filling in'' responses to omitted stimuli occurred only after mice V1 was repeatedly stimulated with sequences of gratings. We proceed similarly in our framework by removing altogether recurrent excitatory connections during sensory learning, and adding them only when training on sequential inputs. Please, note that to keep simple and complex cells' RFs the same throughout the experiments, feedforward excitatory and local lateral inhibitory weights are kept frozen when training PCL+ on the different sequential inputs. We also observed that strong recurrent excitatory inputs perturb sensory learning (Fig.~\ref{fig:suppl1}-b) which further motivates this delayed growth of recurrent excitatory synapses.

\subsection{PCL+ reproduces spatiotemporal sequence recognition and prediction effects observed in V1}

\begin{figure*}[ht]
	\centering
	\includegraphics[width=\textwidth]{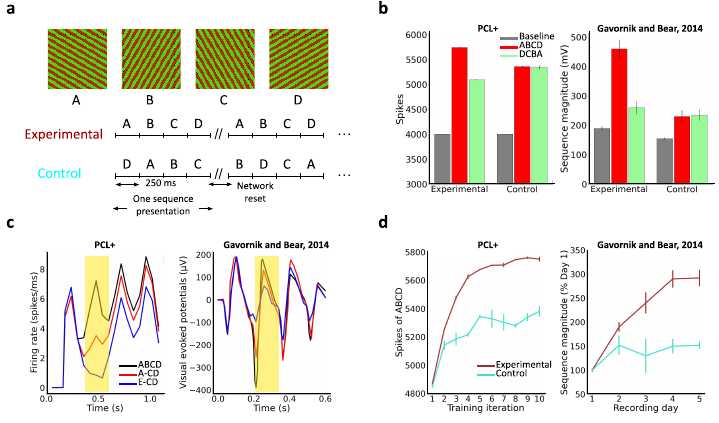}
	\caption[PCL+ network reproduces sequence learning in primary visual cortex.]{\textbf{PCL+ network reproduces sequence learning in primary visual cortex. a,} Experiment structure. A PCL+ network is trained for 10 presentations with a grating sequence ABCD. After each presentation, the state of the network is reset. A control network is also trained for 10 presentations with sequences of randomly permuted elements of ABCD. \textbf{b,} Network responses after training. Left shows the response of the experimental and the control networks and right shows the results observed in mouse V1. The baseline refers to the response of PCL+ with recurrent excitation deactivated. In \cite{gavornik2014learned}, it refers to the response of cells prior to learning or growth of recurrent excitatory synapses. Error bars indicate the standard error across 3 trainings. \textbf{c,} Response of the experimental network to an omitted element of ABCD. Left and right again show the results of PCL+ and those observed in mouse V1. Yellow rectangle indicates the time interval of the second element of the sequences. \textbf{d,} Evolution of network activity during training. Left again shows the response of PCL+ and right shows the results observed in mouse V1. Error bars again indicate the standard error across 3 trainings.}
	\label{fig:figure3}
\end{figure*}

V1 neurons can discriminate as well as predict a given spatiotemporal sequence after being exposed multiple times to the same sequence \cite{gavornik2014learned}. Following the implementation of our delayed recurrent excitatory synapses, we wondered if PCL+'s cells could exhibit this functionality. 

Similarly to Gavornik and Bear's study, we train our PCL+ network on sequences of gratings. Specifically, we consider an experimental network which sees multiple repetitions of a fixed sequence ``ABCD'' and a control network which sees at each iteration a different randomly permuted sequence of these elements (Figure~\ref{fig:figure3}a). Both networks are trained for a total of 10 sequence presentations (adjustable with the learning rate). 

Importantly, in our simulations, we observed that while feedforward excitatory and local lateral inhibitory synapses must be kept frozen to preserve the V1-like receptive fields, distant lateral and top-down inhibitory synapses must however be trained simultaneously with their excitatory counterparts to promote excitation-inhibition balance to avoid runaway activity. 

As in mouse V1, after exposure to sequences of gratings, the experimental PCL+ network responds more effectively to the trained ``ABCD'' sequence than to an untrained ``DCBA'' sequence, while the control PCL+ network does not exhibit a preference for one or the other sequence (Figure~\ref{fig:figure3}b). Also, responses after training are generally stronger than in the baseline PCL network which does not incorporate recurrent excitatory connections.

Additionally, the experimental PCL+ network learns to ``fill in'' missing elements of a sequence. This is tested by exciting the network three types of sequences: the ``ABCD'' sequence on which the PCL+ was trained on, an ``A\_CD'' sequence and an ``E\_CD'' sequence where ``\_'' denotes an omitted (blank) stimulus. We observe in Fig.~\ref{fig:figure3}c that when excited with the ``A\_CD'' sequence, PCL+ exhibits a fill in response during the time interval (colored rectangle) where the element ``B'' was expected to occur, albeit lower than the one observed when ``B'' is actually shown to the network. When excited with the cue ``E'', PCL+ shows little to no response in the next time interval, as the network did not associate ``E'' with a second element ``B''. 

Furthermore, we confirm that this sensitivity to spatiotemporal sequences arises through STDP-learning in our PCL+ network and saturates after a certain number of training iterations for both experimental and control networks (Figure~\ref{fig:figure3}d).

Please, note that while in the biological experiments, the responses of the neurons were recorded using visual evoked potentials (VEP), although we here used spike counts to quantify neuronal responses, in principle, high spike counts (respectively low spike counts) would be analogous to high VEP (respectively high spike counts). Overall, the PCL+ network reproduces different effects observed in \cite{gavornik2014learned}, which suggests that recurrent excitatory connections learning via STDP are a good candidate mechanism to explain these phenomena. 

\subsection{The range of synaptic delays play a role in spatiotemporal sequence learning}

\begin{figure*}[ht]
	\centering
	\includegraphics[]{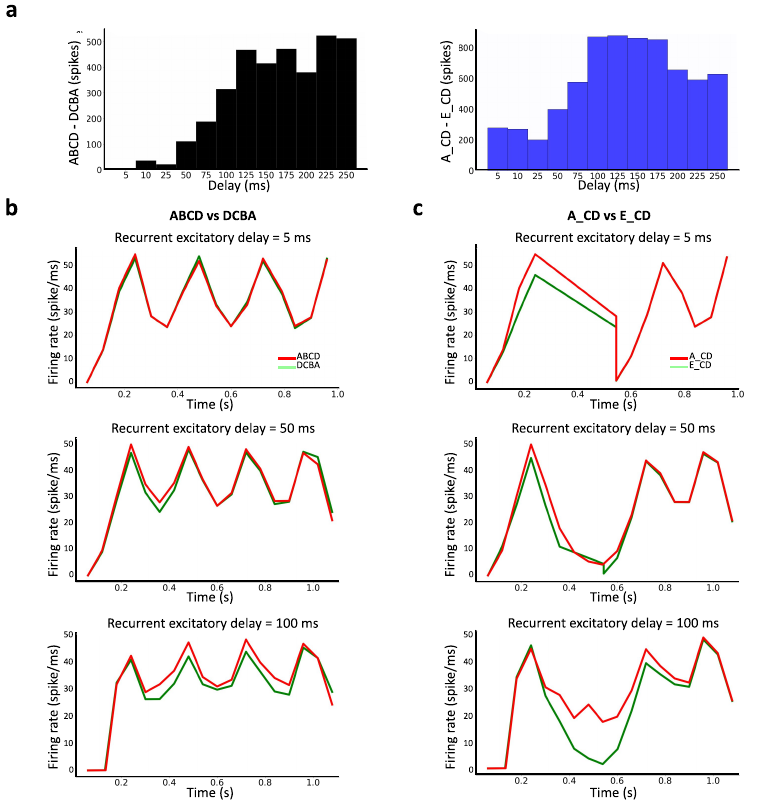}
	\caption[Successful prediction in the PCL+ network requires sufficiently long connection delays.]{\textbf{Successful prediction in the PCL+ network requires sufficiently long connection delays. a,} Difference in the number of spikes generated for sequences ``ABCD'' and ``DCBA'' (left), and for sequences ``A\_CD'' and ``E\_CD'' (right) for different synaptic delays for single simulations. \textbf{b,} Spiking rate over time of the PCL+ network for 3 recurrent excitatory delays (again identical for both distant lateral and top-down excitatory synapses) after excitation with gratings sequences ``ABCD'' (in red) and ``DCBA'' (in green) for a single simulation of the PCL+ network. The gap between PCL+'s response to ``ABCD'' and ``DCBA'' increases with the delay. \textbf{c,} Same as b but for gratings sequences ``A\_CD'' (in red) and ``E\_CD'' in green for a single simulation of the PCL+ network. Again, the gap between PCL+'s response to ``A\_CD'' and ``E\_CD'' increases with the delay.}
	\label{fig:figure4}
\end{figure*}

We next wondered about the impact of synaptic delays on the emergence of these effects in the PCL+ network. To study the effect of these delays, we here replace the uniform distribution of delays with a unique delay for all recurrent excitatory synaptic connections and then proceed to test different delays. 

Our intuition is the following: to bind one sequence element to another in time, delays need to be sufficiently long. 

Our results confirm this hypothesis as short delays (below 100~ms) cannot elicit memory effects (Figure~\ref{fig:figure4}a).  The increased response of the PCL+ network to one particular sequence as well as its ability to fill in omitted inputs thus strongly depend on the synaptic delays used. Figures~\ref{fig:figure4}b-c show examples of the responses, in time, for different delays. Interestingly, although each element of a sequence lasts 250~ms, fill-in responses are already observed for a delay of 100~ms, which we hypothesize to be due from loops of recurrent excitation within the network. 

Essentially, information is able to propagate in time to bind different inputs in the neural representation. We discuss in Sec.~\ref{sec:discussion} how the synaptic delays used in the PCL+ network relate to biological neural processing.

\subsection{PCL+ learns to predict gestures}

\begin{figure*}[ht]
	\centering
	\includegraphics[width = \textwidth]{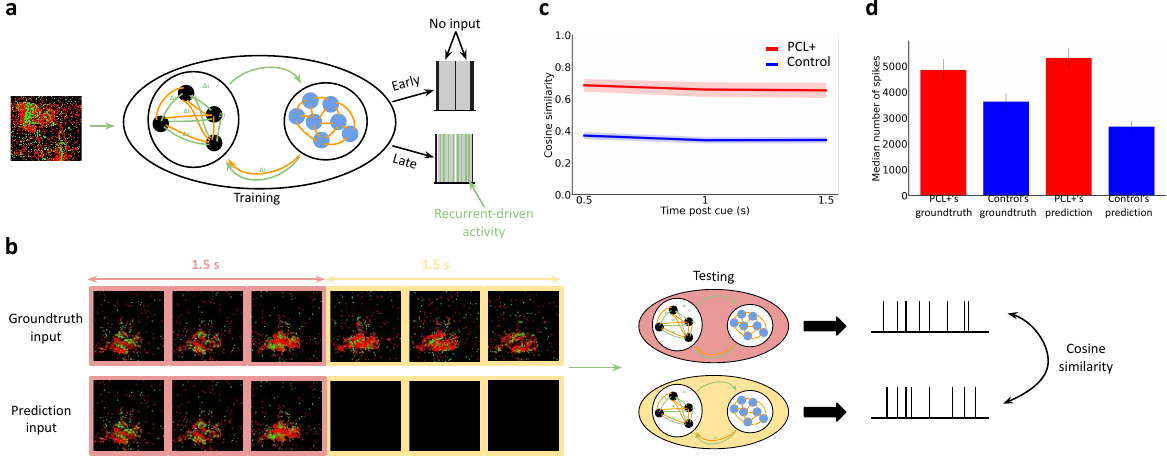}
	\caption{\textbf{Predictive Coding Light+'s recurrent-driven activity predicts neural activity elicited by gestures better than a control network which uses random excitation. a,} Training protocol. The PCL+ network's distant lateral and top-down inhibition/excitation are trained on gestures. At first, with random connectivity, little recurrent-driven activity is elicited. After learning, strong recurrent-driven activity occurs. \textbf{b,} Experimental protocol. After training, the PCL+ network is excited with two input types: a groundtruth in which a 3~s long gesture sample is shown to the network, and a prediction input in which a 1.5~s input is presented to the network and no input is shown for the remaining 1.5~s. The spikes of the PCL+ network to the two types of input are recorded and the cosine similarity between the two spike train descriptors (see Methods) is calculated. \textbf{c,} Cosine similarity between groundtruths and predictions for both the PCL+ network and a control network that uses random excitatory connections (with the same normalization factor as the trained network, see Methods). Results show averages on 3 networks for all gesture samples. \textbf{d,} Number of spikes elicited by PCL+ and the control network when excited with the groundtruth input and the prediction input. }
	\label{fig:figure5}
\end{figure*}

As PCL+ was shown above to be able to fill in missing inputs, we speculated that this could apply not only to sequences of gratings, but also other types of visual sequences. We evaluated this hypothesis with real-life event-based data by using the well-known DVS128 Gesture dataset. We again trained our PCL+ network by taking an already trained PCL network and training only its distant lateral excitation and inhibition as well as the top-down excitation and inhibition. We did this training for 2 epochs using only the first 3.5 seconds of all the samples of the database's training set (Fig.~\ref{fig:figure5}a). 

We subsequently tested this trained network on a time-series forecasting task. Our intuition was as follows: if the recurrent activity can be used to fill in missing inputs, then from a standard gesture elicited neural activity, PCL+ should be able to altogether predict future neural activity. We excited our trained PCL+ network with a cue gesture input that lasted 1.5~s to generate neural activity in the network. We then removed the input and let the recurrent activity be used to predict future spikes and compared this predicted spike train with the groundtruth spike train by calculating the cosine similarity between the two normalized spike trains (Figure~\ref{fig:figure5}b). Note that the groundtruth spike train is the spike train of the PCL+ network when the input gesture persists. To evaluate the quality of this learned prediction, we compare the resulted cosine similarity with a control case in which we use an untrained network that has random recurrent excitatory connections, albeit with the same total synaptic strength as the learned PCL+ network. 

We see that the PCL+'s spiking activity exhibits a higher degree of similarity (almost doubled) to the groundtruth spike train than the control network's spike train, and that the representation only deteriorates minimally with time (Fig.~\ref{fig:figure5}c). Additionally, we note that PCL+ always leads to more activity as the control network (Fig.~\ref{fig:figure5}d), which is also consistent with the results observed with sequences of gratings.

Overall, our results show that PCL+ learns to predict complex spatiotemporal sequences such as gestures. This importantly showcases the remarkable ability of eSTDP and iSTDP to learn to associate different features with each other during learning. 

\section{Discussion}\label{sec:discussion}

We have presented {\em Predictive Coding Light+} (PCL+), an extension of the Predictive Coding Light model \cite{n2025predictive} that uses excitatory and inhibitory spike timing-dependent plasticity (eSTDP/iSTDP) in a recurrent spiking neural network with heterogeneous delays to learn temporal structure from event-based visual data. 
After multiple exposures to a sequence of gratings, neurons of the PCL+ network distinguished between the learned sequence, and an irregular sequence. PCL+ also exhibited the a ``filling-in'' behavior, in which neurons complete a missing input. Our results highlight how heterogeneous synaptic transmission delays allow a spiking neural network to learn a working memory. Additionally, the emergence of this type of neural response to temporally predictable inputs showed the capacity of the PCL+ network to perform associative learning by learning temporal associations. Interestingly, PCL+ also predicted more complex event-based visual sequences than gratings such as gestures. 

While previous works \cite{hartmann2015s, klos2018bridging} also reproduced the results of \cite{gavornik2014learned} by pairing STDP and recurrent excitation, our work uses input from an event-based vision sensor and learns plausible visual receptive fields. Further, in the context of predictive coding, while PCL could generate surround suppression effects seen as hallmarks as predictive codes, PCL+ further adds to these prediction-error like responses with its filling-in properties to signal a mismatch between prediction and input. Other spiking predictive coding works also learned filling-in like properties by reconstructing images, albeit in space rather than in time \cite{ororbia2023spiking, lee2024predictive}. 

While the PCL+ network is not primarily a neurobiological model as it violates Dale's law and uses somewhat {\em ad hoc} activity stabilization mechanisms, it provides some insights into how the interplay of delays, eSTDP and iSTDP could build associative memories in biological spiking neural networks. Interestingly, V1 responses are also thought to be modulated by the hippocampus to learn how to process sequences \cite{finnie2021spatiotemporal}. Our activity stabilization mechanisms could further be viewed as a correlate of the influence of the hippocampus on V1. 


Some of the effects observed in mammalian V1 such as higher spiking activity (and thus a larger energy expenditure) for trained stimuli appear at odds with the theory of predictive coding \cite{westerberg2025hierarchical}. This suggests that a trade-off between the energy efficiency posited by predictive coding, and behavioral requirements must be considered. The PCL+ network can be seen as implementing such a trade-off.

Overall, PCL+ shows how networks of spiking neurons can build a working memory and learn associative structures by combining synaptic delays and biologically plausible spike timing-based local plasticity mechanisms. The PCL+ network uses recurrent excitatory connections with heterogeneous delays, some of which are potentiated through local learning mechanisms. While this seems biologically plausible, it is an open question if directly adapting the delay of particular connections \cite{hammouamri2023learning, grimaldi2023learning} could make learning more efficient.

\section{Acknowledgements}
This work benefited from a French government grant managed by the National Research Agency (ANR) under the ``Investissements d'Avenir" program (PIA) with the following reference: ANR-20-SFRI-0003. This work was also supported by the Deutsche Forschungsgemeinschaft (DFG, German Research Foundation) through the Excellence Cluster EXC3066 ``The Adaptive Mind'' and the priority program SPP 2041, project numbers TR 881/7-2; TR 881/8-2  and priority program SPP 2411, project numbers 520617944; TR 881/11-1 and Research Unit FOR 5368. JT acknowledges support from the Johanna Quandt foundation.

\section{Data and code availability statement}
The source code on which the manuscript is based is available at \url{http://github.com/comsee-research/Predictive-Coding-Light/}. The modified source code and (non-public) datasets used here will be made available upon manuscript's acceptance.

\section{Conflict of interest}
The authors declare no conflict of interest.

\printbibliography

\clearpage
\onecolumn
\centerline{\textbf{\Huge Supplementary material}}

\setcounter{table}{0}
\setcounter{figure}{0}
\renewcommand{\thetable}{S\arabic{table}}
\renewcommand\thefigure{S\arabic{figure}}
\renewcommand{\theHtable}{Supplement.\thetable}
\renewcommand{\theHfigure}{Supplement.\thefigure}

\begin{figure*}[htbp!]
  \centering
  \includegraphics[width=1\textwidth]{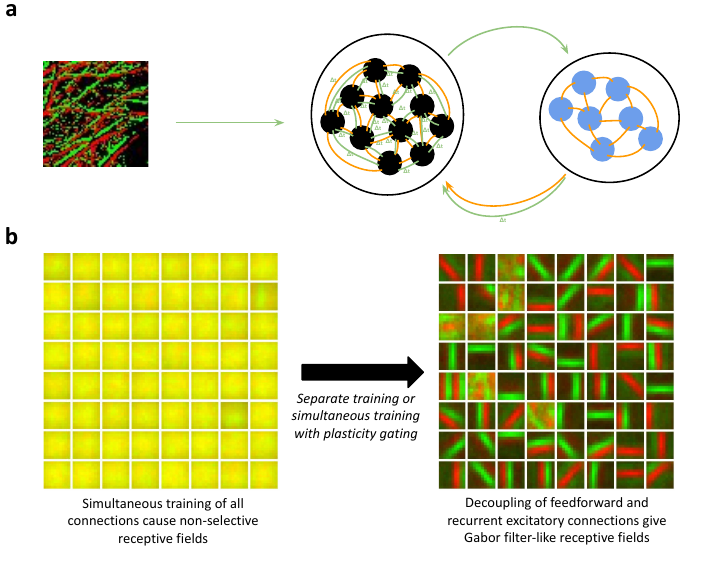}
  \caption{\textbf{Impact of recurrent excitation on sensory learning in a PCL+ network. a,} Simplified network architecture. Dark circles represent simple cells and blue circles represent complex cells. \textbf{b,} Effect of a naive training (left) of simple cells' weights with a PCL+ network, and the training approach (right) used in our simulations.}
  \label{fig:suppl1}
\end{figure*}

\begin{figure*}[htbp!]
  \centering
  \includegraphics[width=0.4\textwidth]{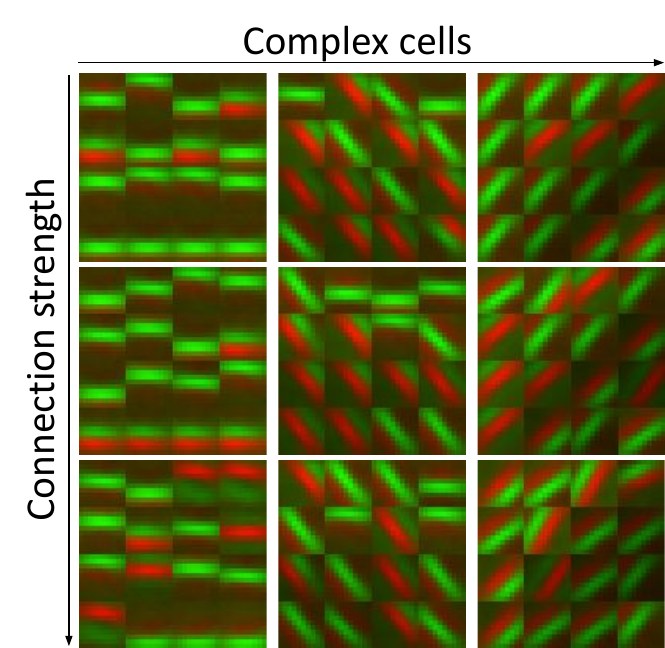}
  \caption{\textbf{Learnt features of example complex cells. The features of complex cells are constructed by visualizing the simple cells to which they have the strongest connectivity to. Rows show features according to the connectivity strength (higher and brighter the better). Columns show different complex cells.}}
  \label{fig:suppl2}
\end{figure*}
\end{document}